\documentclass[conference]{IEEEtran}
\IEEEoverridecommandlockouts

\usepackage{cite}
\usepackage{amsmath,amssymb}
\usepackage{graphicx}
\usepackage{booktabs}
\usepackage{array}
\usepackage{url}
\usepackage{balance}
\usepackage{xcolor}

\newcommand{\COBA}{COBA}
\newcommand{\CAV}{CAVEMAN}

\title{From Caveman to COBA: Coordinated Optimization-Based Attacks on Grid-Edge Smart Inverters in DER-Rich Power Systems}

\author{
\IEEEauthorblockN{Emad Abukhousa, Anna Raymaker, Animesh Chhotaray, Daniel Molzahn, Saman Zonouz, Sakis Meliopoulos}
\IEEEauthorblockA{School of Electrical and Computer Engineering,
Georgia Institute of Technology,
Atlanta, GA, USA\\
\{emadak, araymaker3, achhotaray3, molzahn, szonouz6, sakis.m\}@gatech.edu}
}

\begin{document}
\maketitle

\begin{abstract}
Aggregation platforms and smart-inverter control channels can allow a compromised coordination layer to alter active- and reactive-power commands across a geographically distributed DER fleet. This paper introduces the Coordinated Optimization-Based Attack (\COBA{}), which uses network information and location-specific DER $P/Q$ setpoints to maximize thermal loading on a selected transmission corridor. \COBA{} is compared with low-information \CAV{} baselines limited to a single uniform fleet command: all-off, all-on, maximum reactive absorption, or maximum reactive injection. The Hawaii Synthetic Grid--37 Buses case is evaluated with rooftop-photovoltaic-like capability aggregated at 20 selected 69-kV buses and verified using MATPOWER power flow. Under the case's single-reference-bus balancing convention, \COBA{} overloads B86 (WAIALUA69--WAIANAE69) near 30\% penetration and B51 (COCONUT69--PALM69) near 55\%, while same-target uniform loadings remain lower and uniform commands primarily stress other natural weak lines. The results show that topology, target selection, and command capability determine DER-induced thermal risk beyond aggregate penetration alone.
\end{abstract}
\bstctlcite{BSTcontrol}

\section{Introduction}

Distributed energy resources (DERs) connected through smart inverters are extending distribution-level control actions into the operational boundary of the bulk power system. Aggregators, vendor clouds, and fleet-management platforms can coordinate active-power management and reactive-power support across many devices. Compromise of such a coordination layer can convert cyber access into geographically distributed physical actuation. IEEE Std. 1547-2018 defines DER interconnection and interoperability requirements, while IEEE Std. 1547.3-2023 provides cybersecurity guidance for DER interconnections \cite{ieee1547,ieee15473}. NERC and the U.S. Department of Energy identify DER aggregation and associated control surfaces as emerging bulk-system reliability and security concerns \cite{nerc2022,doe2022}.

Prior work identifies three relevant attack classes. Load-altering attacks, BlackIoT, and MaDIoT demonstrate that coordinated access to grid-connected devices can create bulk-system consequences and that timing and geographic selection matter \cite{mohsenian2011,soltan2018,shekari2022}. However, these studies primarily manipulate demand rather than inverter $P/Q$ setpoints. False-data-injection and load-redistribution attacks use network knowledge to influence state estimation or operator decisions \cite{liu2011,liang2016}. The attack considered in this paper instead directly actuates DER commands. Smart-inverter studies and the SUN:DOWN disclosure show that inverter, monitoring, and cloud control paths can expose active- and reactive-power functions \cite{ahn2024,musleh2024,forescout2025}. Reactive-power manipulation has also been linked to voltage and protection effects at individual prosumer sites \cite{kandasamy2020}. While these results establish a credible command-injection surface, they leave a critical gap: existing work does not quantify when a fixed compromised DER fleet can be coordinated to redirect thermal stress toward a selected transmission corridor.

The key question is how a fixed compromised DER fleet can be coordinated through network-aware, location-specific commands to maximize loading on a selected transmission corridor. Uniform all-on, all-off, and all-reactive commands serve as low-information baselines that reveal corridors naturally sensitive to fleet-wide disturbances, but they do not determine whether the same fleet can make the selected corridor limiting. This defines the screening problem: holding the DER fleet, penetration definition, selected corridor, and AC verification procedure fixed while varying attacker information and command capability.

This paper presents two attacks, the Coordinated Optimization-Based Attack (\COBA{}) and \CAV{}, as adversarial approaches to the DER-screening problem. \CAV{} represents a low-information attacker limited to one uniform fleet command, whereas \COBA{} represents a network-aware attacker that selects location-specific $P/Q$ setpoints using topology, branch ratings, DER locations, capability limits, and an operating point. The comparison determines whether \COBA{} can overload the selected corridor at a lower penetration than under uniform commands or when no uniform command reaches its thermal rating within the evaluated range. This study makes two contributions. First, it formulates a common AC screening framework that compares low-information uniform DER commands with network-aware, location-specific $P/Q$ coordination under the same compromised fleet, penetration definition, target corridor, and verification procedure. Second, it applies the framework to the Hawaii Synthetic Grid--37 Buses case to quantify first-overload thresholds and distinguish naturally stressed corridors from corridors made limiting through coordinated commands.
\newpage

\section{Threat Model and Screening Problem}
\subsection{Compromise and Attacker Profiles}

In this paper, we consider the following threat model. An attacker can compromise the web-based administrator or monitoring portal that manages a fleet of rooftop photovoltaic systems. Such a portal reaches the underlying inverter control interfaces, for example, Modbus active- and reactive-power setpoint registers, so portal access grants command authority \cite{forescout2025}. The compromise is therefore of a coordination layer, and the commands act on real DER devices rather than directly on transmission buses; the transmission-level model represents their aggregate effect at selected buses. Continuous $P/Q$ selection is treated as an upper-bound command capability. Deployed interfaces may instead expose discrete modes, restricted ranges, or rate-limited commands.
\\

The two attacker profiles differ in both information and command capability. \CAV{} has fleet command access but \textit{no} network model and is intended to emulate a cyber-only attacker that is not familiar with the physics of the grid; it applies one uniform rule to every compromised DER. \COBA{} has access to topology, branch ratings, an operating point, DER locations, and capability limits, and issues location-specific commands. Such information could be obtained through insider access, leaked models, compromised utility systems, or reconnaissance. Both profiles send simultaneous fixed active- and reactive-power commands, but neither trips lines, changes relay settings, alters measurements, nor controls non-DER loads.

\subsection{Screening Objective and Model Scope}

The screening objective is to identify the first DER penetration level at which each attacker profile causes the selected corridor to exceed its thermal rating, or to determine that no crossing occurs within the tested penetration range. \COBA{} seeks thermal overload without relying on voltage violations or loss of AC power-flow solvability. Voltage collapse, instability, protection operation, and post-command control actions are outside the present scope.

The compromised rooftop-photovoltaic fleet is represented at the transmission level as aggregate controllable $P/Q$ injections at selected load buses; individual devices and distribution feeders are not modeled explicitly. The detailed aggregation and inverter-capability model is presented in Section~III.

The analysis adopts a single-reference-bus, or slack-bus, AC power-flow convention. Under this convention, all non-reference generators remain at their scheduled active-power outputs, while the generator at the reference bus balances the residual active-power mismatch and the associated change in network losses. An automatic generation control (AGC) response model, in which regulating units share the mismatch through economic participation factors, will be considered in future work.

\section{Attack Formulations}
\subsection{DER Command and Capability Model}
Individual rooftop photovoltaic systems and distribution feeders do not appear explicitly in transmission-level models. Consistent with NERC SPIDERWG guidance \cite{nerc_spiderwg_der_aggregation} that DERs may be represented in aggregate at the bulk-power-system bus for reliability studies, rooftop-photovoltaic-like capability is modeled as controllable aggregate net injection at selected high-load buses. The abstraction preserves location-dependent transmission sensitivity without treating the DER additions as conventional voltage-controlled generators.

\setlength{\abovedisplayskip}{2pt plus 1pt minus 1pt}
\setlength{\belowdisplayskip}{2pt plus 1pt minus 1pt}
For DER bus $k$ at penetration level $p$, the available active power, apparent-power rating, and reactive headroom at full active output are given by
\begin{equation}
P^{\mathrm{av}}_k(p)=\frac{p}{100}P^{\mathrm{load}}_k
\end{equation}

\begin{equation}
S_k^{\max}(p)=\frac{P_k^{\mathrm{av}}(p)}{\mathrm{pf}_{\min}}
\end{equation}

\begin{equation}
Q^{\mathrm{marg}}_k(p)=\sqrt{\left(S_k^{\max}(p)\right)^2-\left(P_k^{\mathrm{av}}(p)\right)^2}
\end{equation}
where $\mathrm{pf}_{\min}$ denotes the minimum power factor at full
available active-power output. The command set for each DER is
\begin{align}
0 &\leq P_k\leq P^{\mathrm{av}}_k
\end{align}
\begin{align}
-S_{k}^{\max} &\leq Q_k\leq S_{k}^{\max}
\end{align}
\begin{equation}
P_k^2+Q_k^2\leq \left(S_{k}^{\max}\right)^2
\end{equation}
The value $Q_k^{\mathrm{marg}}$ characterizes reactive headroom when
$P_k=P_k^{\mathrm{av}}$; \COBA{} may curtail active power and use the
feasible inverter $P/Q$ region. Positive $P_k$ and $Q_k$ denote active-
and reactive-power injection into the network, while negative $Q_k$
denotes reactive-power absorption. The capability model supports
transmission-level adversarial screening but does not represent
site-specific settings, feeder controls, or discrete interface
restrictions.

\subsection{Uniform Low-Information Baselines}
The \CAV{} baselines issue the same rule to every compromised DER. C-OFF commands zero output, C-ON commands full available active power at unity power factor, and the pure-reactive cases convert available inverter capability entirely into reactive absorption or injection:
\begin{align}
\mathrm{C\mbox{-}OFF:}\quad &P_k=0,\quad Q_k=0
\end{align}
\begin{align}
\mathrm{C\mbox{-}ON:}\quad &P_k=P_k^{\mathrm{av}},\quad Q_k=0
\end{align}
\begin{align}
\mathrm{C\mbox{-}QA:}\quad &P_k=0,\quad Q_k=-S_k^{\max}
\end{align}
\begin{align}
\mathrm{C\mbox{-}QI:}\quad &P_k=0,\quad Q_k=+S_k^{\max}
\end{align}
C-QA denotes reactive absorption and C-QI denotes reactive injection, where $Q_k$ is the reactive power injected or absorbed at bus $k$.

\subsection{COBA Formulation}
\COBA{} chooses individualized setpoints for the compromised DER fleet.
\begin{equation}
u=[P_1,Q_1,\ldots,P_{N_D},Q_{N_D}]^T
\end{equation}
For target branch $\ell^\star$ and terminal side $s\in\{f,t\}$, \COBA{} solves
\begin{equation}
\label{eq:coba}
\begin{aligned}
\max_{u,V,\theta,P_r^G,Q^G}\quad
&\frac{(P_{\ell^\star}^{s})^2+(Q_{\ell^\star}^{s})^2}
{(S_{\ell^\star}^{\mathrm{rate}})^2} \\
\mathrm{s.t.}\quad
&u\in\mathcal{U}(p) \\
&P_i^G-P_i^D+P_i^{\mathrm{DER}}(u)=F_i^P(V,\theta),
\quad i\in\mathcal{N} \\
&Q_i^G-Q_i^D+Q_i^{\mathrm{DER}}(u)=F_i^Q(V,\theta),
\quad i\in\mathcal{N} \\
&P_i^G=P_i^{G,0},\quad i\in\mathcal{G}\setminus\{r\},
\qquad P_r^G\in\mathbb{R} \\
&Q_i^G\in\mathbb{R},\quad i\in\mathcal{G} \\
&0.88\leq |V_i|\leq 1.10\ \mathrm{p.u.},
\quad i\in\mathcal{N}
\end{aligned}
\end{equation}
where $\mathcal{U}(p)$ is the inverter command set in (4)--(6); $\mathcal{N}$ and $\mathcal{G}$ are the bus and conventional-generator sets; $r$ denotes the reference generator; $P_i^{G,0}$ is the scheduled active output; $\mathcal{D}_i$ is the compromised DER aggregation set at bus $i$; $N_D$ is the number of compromised DER aggregation buses; $P_i^G,Q_i^G$ and $P_i^D,Q_i^D$ are conventional generation and demand; $V_i,\theta_i$ are bus voltage magnitude and angle; $\theta_{ij}=\theta_i-\theta_j$; and
\begin{equation}
P_i^{\mathrm{DER}}(u)=\sum_{k\in\mathcal{D}_i}P_k
\end{equation}
\begin{equation}
Q_i^{\mathrm{DER}}(u)=\sum_{k\in\mathcal{D}_i}Q_k
\end{equation}
The explicit active-power constraints in \eqref{eq:coba} fix every non-reference generator at its base schedule and leave $P_r^G$ to absorb the DER-induced mismatch. The domain $Q_i^G\in\mathbb{R}$ makes generator reactive capability limits and voltage-controlled to constant-power (PV/PQ) bus switching relaxed assumptions in the present screen. The voltage bounds constrain the screen to in-band thermal-overload solutions rather than imposing a universal restriction on attacker behavior; protection tripping, instability, and voltage collapse remain outside scope. The AC injection functions are
\begin{equation}
F_i^P=|V_i|\sum_j |V_j|(G_{ij}\cos\theta_{ij}+B_{ij}\sin\theta_{ij})
\end{equation}
\begin{equation}
F_i^Q=|V_i|\sum_j |V_j|(G_{ij}\sin\theta_{ij}-B_{ij}\cos\theta_{ij})
\end{equation}
Branch loading accounts for line losses by taking the larger terminal apparent-power magnitude:
\begin{equation}
L_\ell=100\frac{\max(|S_\ell^f|,|S_\ell^t|)}{S_\ell^{\mathrm{rate}}}\label{eq:loading}
\end{equation}
\setlength{\abovedisplayskip}{3pt plus 1pt minus 1pt}
\setlength{\belowdisplayskip}{3pt plus 1pt minus 1pt}
Here, $G_{ij}$ and $B_{ij}$ are bus-admittance elements, and $P_{\ell^\star}^{s},Q_{\ell^\star}^{s}$ are target-branch terminal flows. The side-specific objective is a proxy for \eqref{eq:loading}; the two
terminal magnitudes differ because of line losses, and the larger terminal value defines the branch loading.

\section{Test Case and Study Setup}
\subsection{Reference Network}

The base network is the Hawaii Synthetic Grid--37 Buses case from the Texas A\&M Electric Grid Test Case Repository \cite{hawaii37}. This 37-bus synthetic network covers the geographic footprint of the Hawaiian island of Oahu and serves 1100 MW of peak load. The footprint is selected because Hawaii has among the highest rooftop-photovoltaic penetration in the United States, making it a representative setting for a DER-rich screening study. The publicly distributed case file is used as the base network model, with its original conventional-generation and branch data retained. No protection or dynamic models are added. The case contains 37 buses, 89 branches, and a 100-MVA base.
\begin{table}[t]
\centering
\caption{DER aggregation buses used in the study}
\label{tab:derbuses}
\scriptsize
\begin{tabular}{r l r@{\hspace{0.30cm}}r l r}
\toprule
Bus & Name & MW & Bus & Name & MW\\
\midrule
26 & EWA BCH69   & 75.28 & 30 & WAHIAWA69   & 49.03\\
11 & MAHALO69    & 65.55 & 19 & PEARL CTY69 & 48.60\\
8  & KANEOHE69   & 65.10 & 21 & AIEA69      & 48.34\\
20 & MILILANI69  & 64.95 & 24 & KAPOLEI69   & 46.58\\
15 & KAILUA69    & 60.90 & 10 & TURTLE69    & 37.76\\
2  & ALOHA69     & 60.70 & 7  & SURF69      & 37.01\\
3  & FLOWER69    & 59.39 & 18 & VOLCANO69   & 36.32\\
13 & COCONUT69   & 59.24 & 6  & HONOLULU69  & 27.46\\
33 & WAIANAE69   & 58.22 & 16 & PALM69      & 23.83\\
12 & LYCHEE69    & 54.01 & 4  & WAVE69      & 22.47\\
\bottomrule
\end{tabular}
\end{table}

\subsection{DER Aggregation and Target Corridors}

The 20 aggregation points are the highest-load eligible buses after excluding the reference bus and generator-only locations ($N_D=20$); one penetration percentage is applied to all selected points, so available DER power scales with local load.

The base-case AC power flow defines the starting operating point, with
WAIPAHU69 as the reference bus. Aggregate DER capability is then assigned to the 20 buses in Table~\ref{tab:derbuses} before target selection. Base-case inspection and preliminary stress screening identify two corridors with different starting conditions. B86, WAIALUA69--WAIANAE69, has a 38.0-MVA Rate A and a verified base flow of 27.09 MVA, or 71.29\% of Rate A. B51, COCONUT69--PALM69, has a 52.0-MVA Rate A and a lower base flow of 18.34 MVA, or 35.26\%. The pair therefore tests an already stressed low-rating corridor and a less naturally stressed corridor under the same fixed DER map. The inverter capability model uses $\mathrm{pf}_{\min}=0.85$.

\subsection{Study Procedure}
The penetration sweep spans 0\% to 100\% in 5-percentage-point
increments. At each level, four \CAV{} command vectors and one optimized
\COBA{} command vector are applied as fixed net injections. Recorded
outputs include branch loadings, target loading, system-maximum branch
identity, bus-voltage extremes, and feasibility status. The \COBA{} optimization is solved in YALMIP with sequential quadratic programming (SQP) through fmincon. Each solve is warm-started from a converged AC operating point and uses $10^{-6}$ constraint and optimality tolerances with a $10^{-8}$ step tolerance. Each resulting command is replayed as a fixed net injection and independently verified using MATPOWER AC power flow, so reported loadings are verified values rather than raw optimizer objectives.

\section{Results}

\begin{figure}[!t]
\centering
\includegraphics[width=\linewidth]{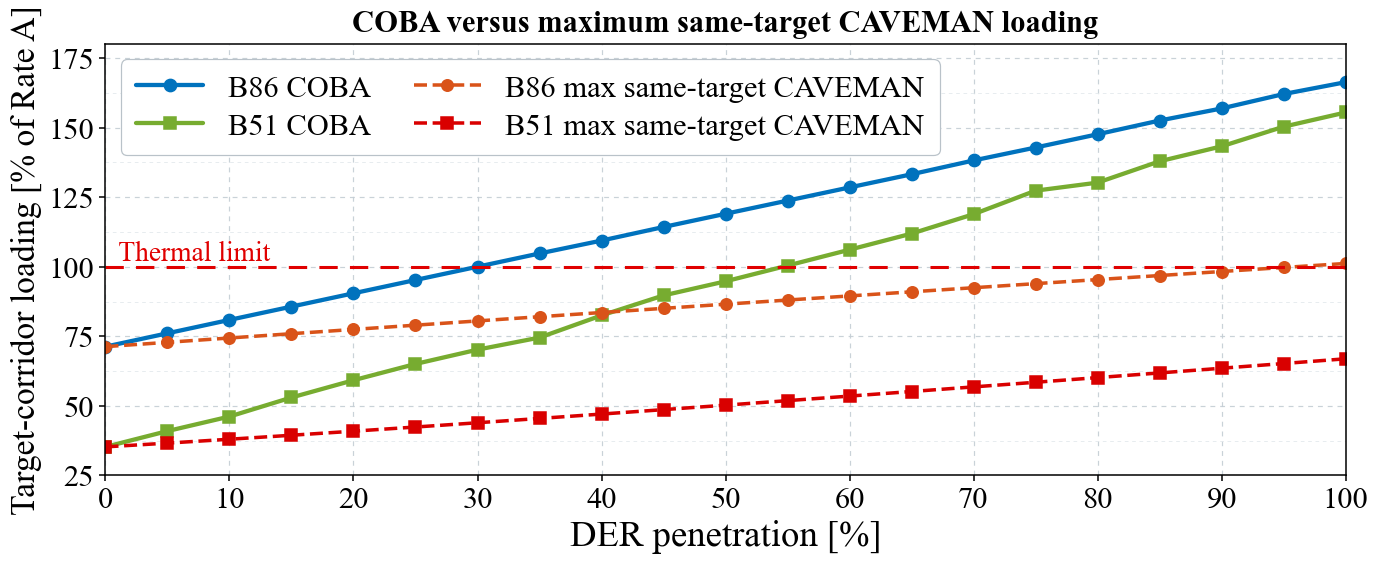}
\caption{\COBA{} versus the strongest same-target \CAV{} baseline. B86 crosses its thermal rating near 30\% penetration, while B51 crosses near 55\%.}
\label{fig:corridors}
\end{figure}

\begin{table}[t]
\centering
\caption{First-overload thresholds}
\label{tab:thresholds}
\scriptsize
\setlength{\tabcolsep}{2.2pt}
\begin{tabular}{c c c c c c}
\toprule
Target & \shortstack{Rate A\\{}[MVA]} & \shortstack{Threshold\\$p$ [\%]} & \shortstack{Strongest\\uniform} & \shortstack{\COBA{} at $p$\\{}[\%]} & \shortstack{Uniform at $p$\\{}[\%]}\\
\midrule
B86 & 38.0 & 30 & C-ON & 100.1 & 81.2\\
B51 & 52.0 & 55 & C-QI & 103.8 & 55.4\\
\bottomrule
\end{tabular}
\end{table}

\begin{figure}[t]
\centering
\includegraphics[width=\linewidth]{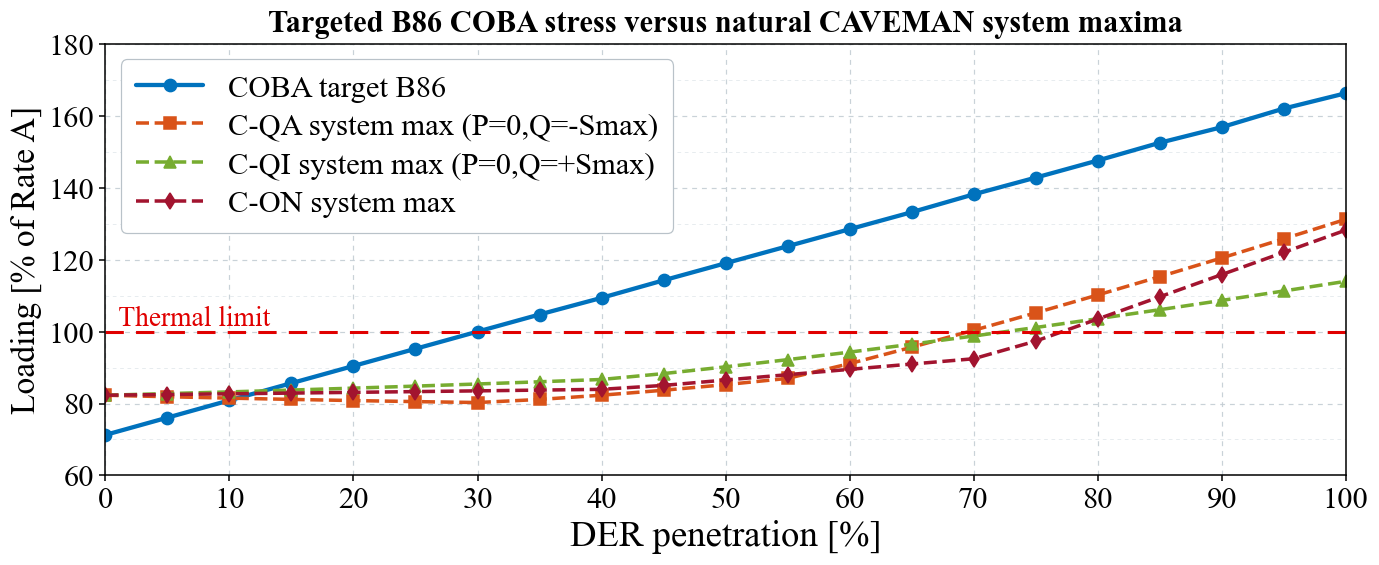}
\caption{Targeted B86 loading compared with system-wide maxima under uniform commands.}
\label{fig:b86natural}
\end{figure}

\subsection{Target-Corridor Overload Thresholds}
Two target corridors are used to avoid a single-line result: B86 (WAIALUA69--WAIANAE69) and B51 (COCONUT69--PALM69). Fig.~\ref{fig:corridors} shows the full penetration sweep, and Table~\ref{tab:thresholds} summarizes the first overload thresholds. \COBA{} overloads B86 at approximately 30\% penetration and B51 at approximately 55\%. The strongest same-target \CAV{} case is C-ON for B86 and C-QI for B51, demonstrating that different corridors react differently to uniform command types.

The two corridors show different regimes. The strongest uniform command overloads B86 only near full penetration (103.3\% versus 166.7\% for \COBA{}), so coordination lowers the crossing to 30\%. No uniform command reaches the B51 rating within the sweep (73.5\% versus 162.3\% for \COBA{}), so coordination makes B51 limiting even though no uniform command overloads it within the sweep.

\subsection{Targeted Stress versus Natural System Maxima}
Fig.~\ref{fig:b86natural} compares targeted B86 stress with the system-wide maximum branch loading under C-QA, C-QI, and C-ON at each penetration. The \CAV{} curves identify the branches that naturally become most stressed under each uniform disturbance, whereas \COBA{} deliberately makes B86 limiting and crosses its rating at 30\%.

\begin{figure}[!t]
\centering
\includegraphics[width=\linewidth]{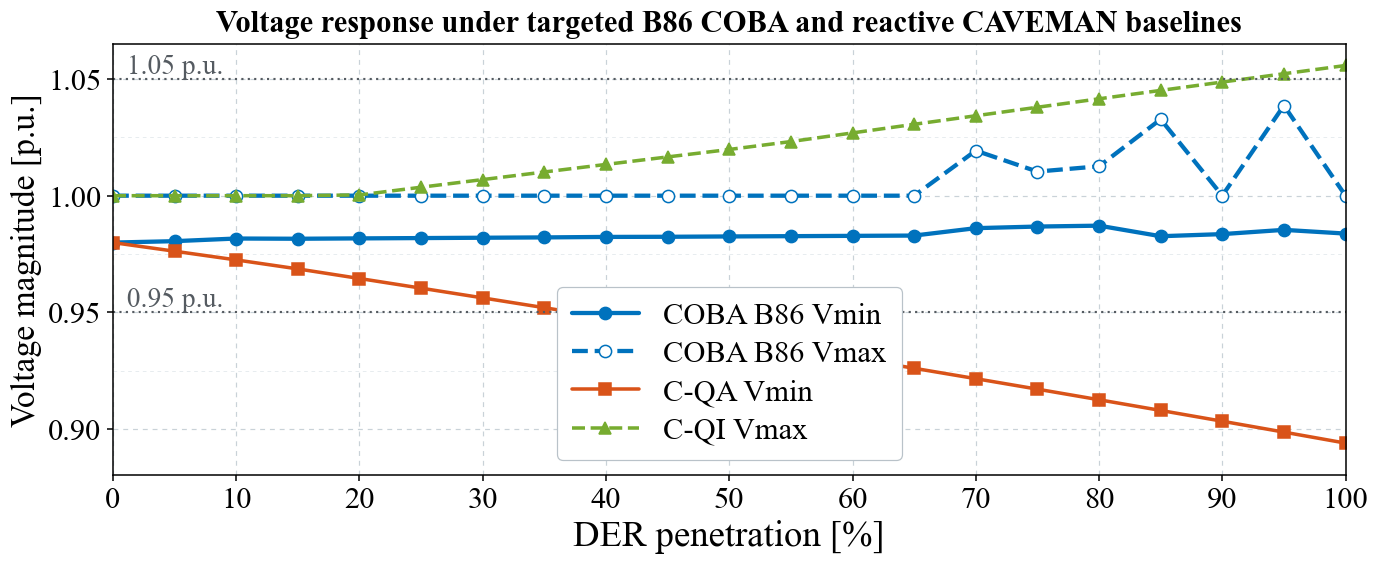}
\caption{Voltage response under \COBA{} and uniform reactive baselines.}
\label{fig:voltage}
\end{figure}

\subsection{Voltage Feasibility}
Fig.~\ref{fig:voltage} shows that the first B86 overload is not the result of a voltage violation. Uniform reactive absorption depresses the minimum voltage, while uniform reactive injection raises the maximum voltage. Thermal-targeted \COBA{} instead coordinates $P/Q$ commands while keeping voltages within the selected envelope near the first crossing. At 30\% penetration, the verified B86 solution has $V_{\min}=0.989$ and $V_{\max}=1.007$ p.u.

\subsection{Target-Specific Attack Strategies}
\begin{figure*}[t]
\centering
\includegraphics[width=0.49\textwidth]{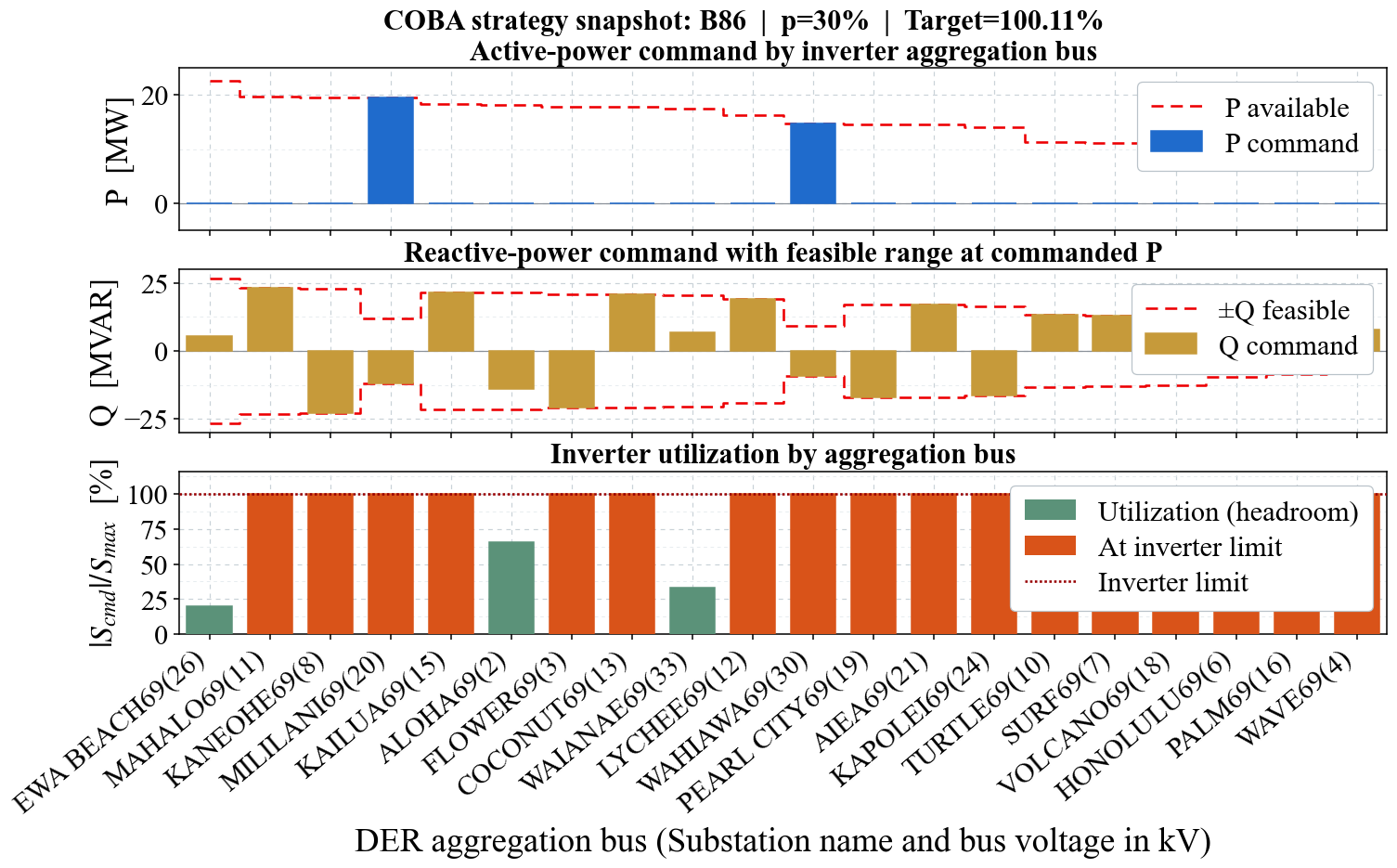}\hfill
\includegraphics[width=0.49\textwidth]{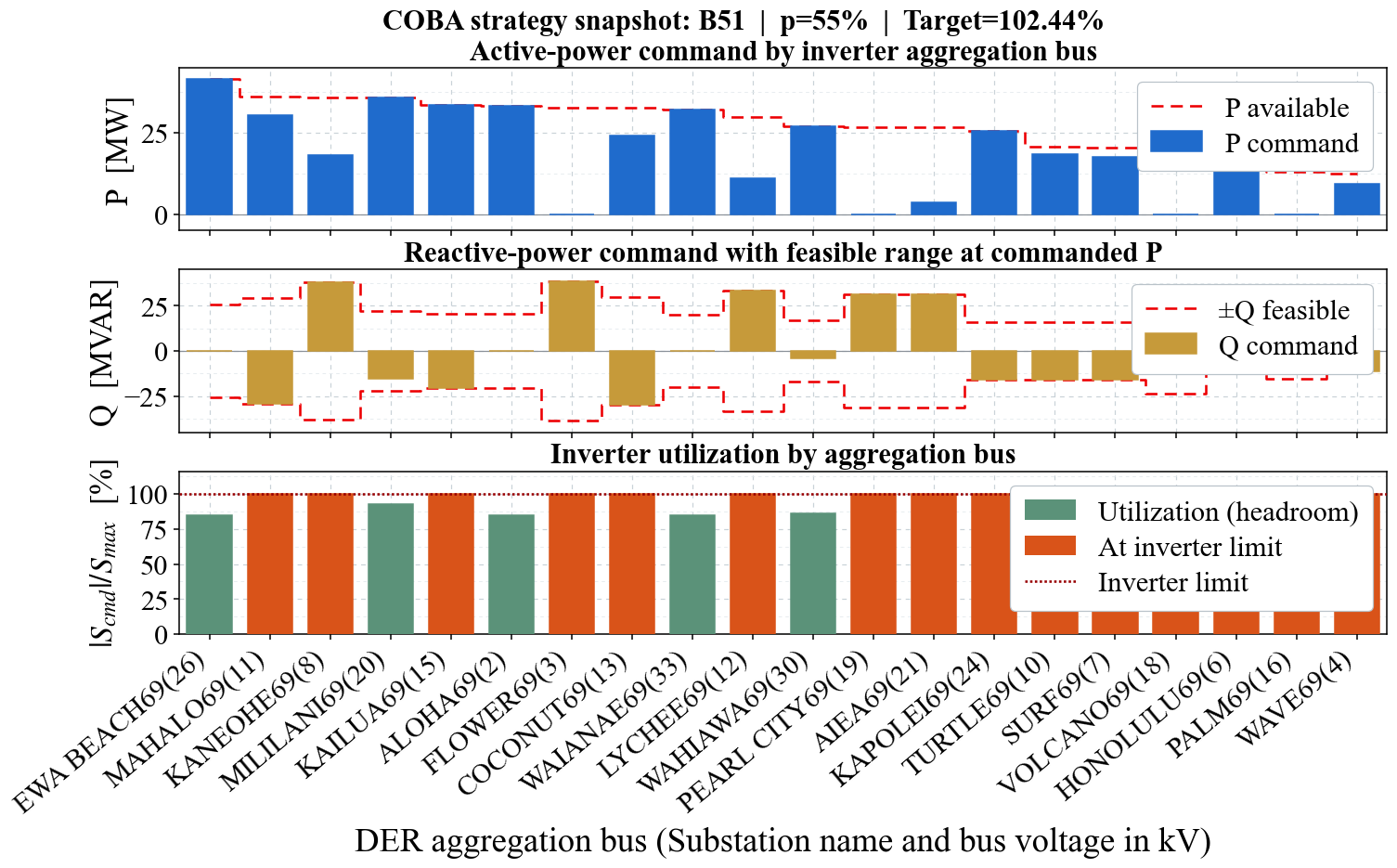}
\caption{Representative \COBA{} strategies near the first overload thresholds. The nonuniform $P/Q$ patterns differ by target. Bus labels use substation-kV(bus no.); ALOHA69(2), for example, denotes MATPOWER bus 2 at the 69-kV ALOHA substation.}
\label{fig:strategy}
\end{figure*}

Fig.~\ref{fig:strategy} shows spatially heterogeneous command patterns near the first overload thresholds. The B86 and B51 strategies use different active- and reactive-power combinations, confirming that \COBA{} redirects stress through location-specific sensitivity rather than uniform fleet saturation. Small or negligible commands at several buses indicate that capability at weakly coupled locations contributes little to the selected corridor.

\section{Discussion and Limitations}\label{sec:discussion}

The results support three conclusions. First, DER penetration alone is not a sufficient security metric because one fleet can produce different overload thresholds across corridors. Second, uniform commands expose natural weak lines but are not conservative for every selected target. Third, branch ratings, DER locations, and command directions determine which inverter capabilities affect each corridor. The reported thresholds are specific to the Hawaii Synthetic Grid--37 Buses case, including its branch ratings, operating point, DER aggregation map, and single-reference-bus convention. Accordingly, the main contribution is a target-aware screening method rather than universal penetration limits. The relevant cyber-physical vulnerability lies in the coordination layer, which translates cyber access into distributed physical actuation. Prior work showed that reactive-power manipulation can disconnect individual prosumer sites \cite{kandasamy2020}; the present results extend this concern to transmission-corridor thermal stress caused by fleet-level $P/Q$ coordination.
\\

Several limitations bound these thresholds. First, all active-power mismatch is assigned to a single reference generator. Future work will replace this assumption with an AGC model in which scheduled units remain at economic dispatch and regulating units share the imbalance through participation factors. Second, generator reactive-power limits and PV/PQ switching are relaxed; enforcing capability curves will clarify their effect on attack feasibility. Third, the steady-state model omits frequency response, protection, redispatch, and transient stability, so dynamic simulation is needed to determine whether overloads persist. Fourth, photovoltaics and feeders enter as aggregate bus injections; feeder-coupled models will add voltage limits, losses, and local controls. Fifth, the full-compromise, full-information attacker with continuous $P/Q$ command authority is an upper bound, so partial-information, discrete-command, and active-power-only cases remain open. Finally, \COBA{} is nonconvex and therefore yields local rather than globally certified solutions; multi-start search, additional networks, optimizer-to-verifier comparisons, and communication-layer studies of portal access and mitigation will be used to assess robustness.

\section{Conclusion}
\vspace{-2pt}

This paper introduces \COBA{}, a network-aware attack executed through a compromised DER coordination layer that controls distributed devices such as rooftop photovoltaic systems. \COBA{} coordinates location-specific DER commands to maximize loading on a selected transmission corridor and is evaluated against low-information uniform fleet commands under a common AC screening procedure. On the Hawaii Synthetic Grid--37 Buses case, \COBA{} overloads corridor B86 near 30\% penetration and B51 near 55\%, while the strongest same-target \CAV{} baselines produce lower loading. The results distinguish natural weak lines exposed by uniform disturbances from selected corridors made limiting through coordinated commands. The reported thresholds are conditional on the case's single-reference-bus convention and represent first-stage screening outcomes. AGC response, generator reactive limits, PV/PQ switching, and dynamic protection remain necessary for more complete operational thresholds.

\bibliographystyle{IEEEtran}
\bibliography{references}

\end{document}